\documentclass[prl,twocolumn,showpacs,amsmath,amssymb]{revtex4}
\usepackage{amsmath}
\usepackage{graphicx}
\usepackage{mathrsfs}
\usepackage{dcolumn}
\usepackage{bm}

\begin{document}


\title{Dissipative Quantum Metrology with Spin Cat States}

\author{Jiahao Huang}
\author{Xizhou Qin}
\author{Honghua Zhong}
\author{Yongguan Ke}
\author{Chaohong Lee}
 \altaffiliation{Emails: chleecn@gmail.com; lichaoh2@mail.sysu.edu.cn}

\affiliation{State Key Laboratory of Optoelectronic Materials and Technologies, School of Physics and Engineering, Sun Yat-Sen University, Guangzhou 510275, China}

\date{\today}

\begin{abstract}
The maximally entangled states are excellent candidates for achieving Heisenberg-limited measurements in ideal quantum metrology, however, they are fragile against dissipation such as particle losses and their achievable precisions may become even worse than the standard quantum limit (SQL).
Here we present a robust high-precision measurement scheme via spin cat states (a kind of non-Gaussian entangled states in superposition of two spin coherent states) in the presence of particle losses.
The input spin cat states are of excellent robustness against particle losses and their achievable precisions may still beat the SQL.
For realistic measurements based upon our scheme, comparing with the population measurement, the parity measurement is more suitable for yielding higher precisions.
In phase measurement with realistic dissipative systems of bosons, our scheme provides a robust and realizable way to achieve high-precision measurements beyond the SQL.
\end{abstract}


\maketitle

Precision metrology and parameter estimation are of great significance in both fundamental sciences and practical technologies. Quantum metrology aims to improve estimation precision via quantum strategy~\cite{Giovannetti2004,Giovannetti2011,LeeBookChapter2014}.
The estimation precision via separable states of $N$ particles is bounded by the standard quantum limit (SQL), which scales as $1/\sqrt{N}$.
The estimation precision can be enhanced by multi-particle quantum correlations, such as quantum entanglement~\cite{Giovannetti2004,Giovannetti2011,LeeBookChapter2014} and quantum discord~\cite{Georgescu2014}.
In particular, by employing maximally entangled states [Greenberger-Horne-Zeilinger (GHZ) states and NOON states], the estimation precision can be improved to the Heisenberg limit (HL)~\cite{Giovannetti2006,Lee2006,Boixo2008, Pezze2009, Liu2011}, which scales as $1/N$.
Entangled atoms could enhance clock accuracy, entangled photons could enhance imaging resolution, and entangled spins could enhance field sensitivity.
The principles of quantum metrology have been extensively used to design practical quantum devices, such as, atomic clocks~\cite{Swallows25022011,Martin09082013,PhysRevLett.92.230801}, gravitational wave detectors~\cite{PhysRevLett.110.093602,PhysRevD.65.022002}, and magnetic field sensors~\cite{PhysRevLett.104.133601,PhysRevA.82.022330,Ng2013}.
In recent, by employing spin squeezed states of Bose condensed atoms, phase sensitivity has been enhanced beyond the SQL~\cite{Gross2010,Riedel2010,Grond2011,Lucke2011,Berrada2013,2014arXiv1405.6022M,UedaPRA}.
Furthermore, by employing non-Gaussian entangled states of Bose condensed atoms, phase sensitivity can also been enhanced beyond the SQL in the absence of spin squeezing~\cite{Strobel2014}.

In realistic experiments, decoherence inevitably exists in the process of signal accumulation.
Most entangled states are sensitive to decoherence and become fragile against particle losses.
In particular, the maximally entangled states are extremely fragile against particle losses and the optimal precision may even be worse than the SQL~\cite{PhysRevLett.102.040403,PhysRevA.80.013825,Joo2011}.
In the quantum metrology via Bose condensed atoms or photons~\cite{Gross2010,Riedel2010,Grond2011,Lucke2011,Berrada2013,2014arXiv1405.6022M}, a typical kind of decoherence is particle losses~\cite{PhysRevA.83.043620,Ockeloen2013,Pawlowski2013,PhysRevLett.100.210401,Strobel2014}.
Therefore, it is a great challenge to find experimentally available states which may achieve high precision and meanwhile are robust against particle losses~\cite{Escher2011,Alipour2014}.
Naturally, two important questions arise: (1) how the particle losses during the signal accumulation process affect the estimation precision? and (2) how to use achievable and robust entangled states to accomplish optimal parameter estimation under particle losses?

In this Letter, we present a robust high-precision phase measurement scheme via quantum interferometry with spin cat states under particle losses.
The phase accumulation process under particle losses is described by a Markovian master equation.
We calculate the ultimate estimation precision for various spin cat states under particle losses.
We find that the spin cat states are robust against particle losses and may still achieve high precision beyond the SQL.
Furthermore, by comparing the optimal precisions achieved by the parity measurement and the population measurement, we find that the parity measurement is more suitable for accomplishing dissipative quantum metrology beyond the SQL.
By using currently available techniques of Bose condensed atoms, spin cat states can be prepared via the Kerr nonlinearity due to atomic collisions~\cite{Strobel2014,Ho2004,Lau2014}, and phase information can be extracted by parity/population measurement via counting atoms~\cite{Hume2013}.

\begin{figure}[htb]
\centering
\includegraphics[width=\columnwidth]{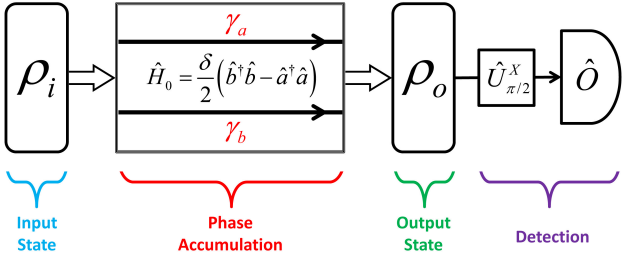}
\caption{(Color online) Schematic diagram for the phase measurement process in the presence of particle losses. $\gamma_a$ and $\gamma_b$ indicate the damping rates in mode $a$ and mode $b$. Here, the input state is prepared to probe and the phase accumulation is induced by a free evolution. The phase information is extracted by applying a $\pi/2$ pulse on the output state and then measuring a certain observable.}
\label{Fig1}
\end{figure}

In Fig.~\ref{Fig1}, we show the schematic diagram of our phase estimation procedure under particle losses.
First, the system is prepared in a desired input state.
Then the input state evolves under the action of the quantity to be measured and then accumulates an unknown phase $\phi$, which is determined by the energy difference $\delta$ and the evolution duration $T$.
Finally, to extract the accumulated phase $\phi$, a proper measurement of the output state is implemented.
Usually, the preparation and the detection can be accomplished in a short period of time.
Therefore, for simplicity, we only consider the dissipation in the phase accumulation process.
Taking into account the dissipation effects on phase accumulation, we will search achievable optimal spin cat states for implementing dissipative quantum metrology.

\begin{figure}[htb]
\centering
\includegraphics[width=\columnwidth]{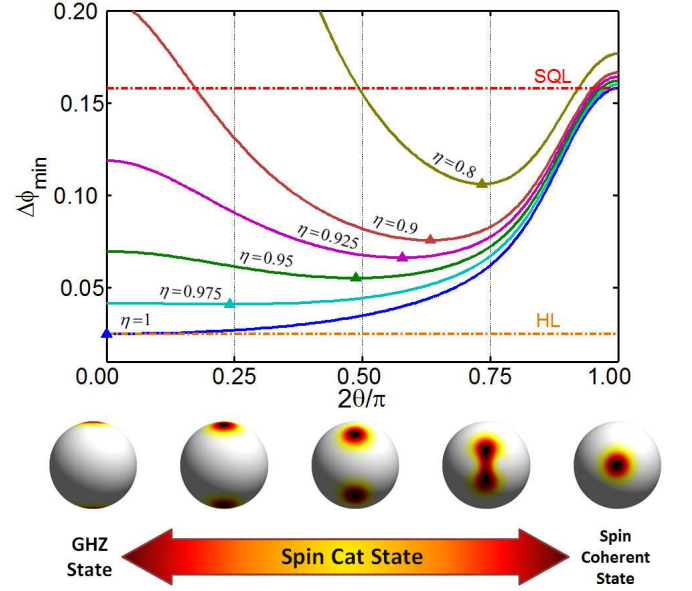}
\caption{(Color online) Top: Estimation precisions of spin cat states versus $\theta$ for different values of $\eta$.
Bottom: Husimi distributions for the input spin cat states.
The triangles denote the best optimal estimation precisions.
The red and orange dash-dotted lines indicate the SQL and HL, respectively.
Here, the initial total particle number $N=40$ and the parameter $\eta=e^{-\gamma T}$ with the damping rate $\gamma$ and the phase accumulation time $T$.}
\label{Fig2}
\end{figure}

The spin cat states are in superpositions of multiple spin coherent states~\cite{Agarwal1997,Gerry1997,Wang2000}. To implement phase measurement, we consider spin cat states in the form of
\begin{equation}\label{CAT}
    |\Psi(\theta, \varphi)\rangle_{\textrm{CAT}}=\mathcal{N}_{C}(|\theta,\varphi\rangle + |\pi-\theta,\varphi\rangle),
\end{equation}
where $\mathcal{N}_{C}$ denoting the normalization factor and $\left|\theta,\varphi\right\rangle$ being the spin coherent state (SCS)
\begin{equation}\label{SCS}
\left|\theta,\varphi\right\rangle =\left[\sin(\frac{\theta}{2})e^{-i\varphi/2} a^{\dagger} +\cos(\frac{\theta}{2})e^{i\varphi/2} b^{\dagger}\right]^N \left|vac\right\rangle.
\end{equation}
Here, the two SCS's have the same azimuthal angle $\varphi$, $N=a^{\dagger}a+b^{\dagger}b$ is the total particle number, $|vac\rangle$ denotes the vacuum state of no any particles, and $\{a^{\dagger}, b^{\dagger}\}$ are bosonic creation operators.
In the Dicke basis $\{\left|J,m\right\rangle\}$, the SCS reads as,
\begin{equation}\label{SCS_Dicke}
    \left|\theta,\varphi\right\rangle=\sum^{J}_{m=-J} c_m(\theta)e^{-i(J+m)\varphi}\left|J,m\right\rangle,
\end{equation}
with $c_m(\theta) =\sqrt{\frac{(2J)!}{(J+m)!(J-m)!}}\cos^{J-m}\left({\theta \over 2}\right)\sin^{J+m}\left({\theta \over 2}\right)$ and $J={N \over 2}$.
Without loss of generality, we assume the azimuthal angle $\varphi=0$ and the initial total particle number $N=40$.
Below we abbreviate $\left|\Psi(\theta, \varphi=0)\right\rangle_{\textrm{CAT}}$ to $\left|\Psi(\theta)\right\rangle_{\textrm{CAT}}$.
For $\theta=0$, $\left|\Psi(\theta)\right\rangle_{\textrm{CAT}}$ is the so-called NOON state~\cite{Sanders2014,Lee2006}, which is a maximally entangled state (GHZ state) in superposition of all particles in mode $a$ and all particles in mode $b$.
For $\theta=\pi/2$, $\left|\Psi(\theta)\right\rangle_{\textrm{CAT}}$ is actually the SCS with no entanglement.
In the region of $0\le\theta\le\pi/2$, the degree of entanglement decreases with the polar angle $\theta$.
In the bottom of Fig.~\ref{Fig2}, we show Husimi distributions for spin cat states with different $\theta$.
For $0\le\theta<\pi/2$, there are two peaks in each Husimi distribution.
In particular, for modest $\theta$, the two peaks are entirely separated.
It has been theoretically demonstrated that spin cat states can be prepared via nonlinear Kerr effects~\cite{Sanders1989,Ho2004,Lee2006,PhysRevLett.102.070401} and cavity-QED state reduction~\cite{Gerry1998,Dooley2013,McConnell2013}.
In experiments, spin cat states have been generated in atomic Bose-Einstein condensates via nonlinear dynamical evolution~\cite{Lau2014} and in thermal atoms via confined quantum Zeno dynamics~\cite{Signoles2014}.

The phase accumulation is governed by the Hamiltonian $\hat{H}_{0}=\delta (\hat{b}^{\dagger} \hat{b}-\hat{a}^{\dagger} \hat{a})/2=\delta \hat{J}_{z}$ with $\delta$ being the inter-mode energy difference. Without loss of generality, we assume $\delta=1$.
The input state $\left|\Psi(\theta)\right\rangle_{\textrm{CAT}}$ will evolve into a phase-dependent state $|\Psi(\theta,\phi)\rangle =\exp(-i\hat{H}_{0}T)\left|\Psi(\theta)\right\rangle_{\textrm{CAT}} =\exp(-i\phi \hat{J}_{z}) \left|\Psi(\theta)\right\rangle_{\textrm{CAT}}$, where $\phi=\delta T$ is the relative phase.
In phase accumulation, the Bose condensed atoms may collide with the residual thermal atoms and then be kicked out of the trap. This dissipative process is well described by one-body loss~\cite{Ockeloen2013,PhysRevLett.102.040403}.
Therefore the phase accumulation obeys a Markovian master equation~\cite{PhysRevA.58.R50,pawlowski2010,Ng2013},
\begin{eqnarray}\label{MasterEq}
\frac{\partial \rho}{\partial t} = &-&i \left[ H_{0},\rho \right] +\gamma_{a}\left(\hat{a}\rho\hat{a}^{\dagger} -\frac{1}{2}\hat{a}^{\dagger}\hat{a}\rho -\frac{1}{2}\rho\hat{a}^{\dagger}\hat{a}\right)\nonumber\\
&+&\gamma_{b}\left(\hat{b}\rho\hat{b}^{\dagger} -\frac{1}{2}\hat{b}^{\dagger}\hat{b}\rho -\frac{1}{2}\rho\hat{b}^{\dagger}\hat{b}\right),
\end{eqnarray}
where $\rho$ is the reduced density matrix, and $\gamma_{a,b}$ are the damping rates.
In our calculation, we assume $\gamma_a=\gamma_b=\gamma$ and therefore the amount of particle losses is given as $N(1-\eta)$ with $\eta=e^{-\gamma T}$.
The master equation \eqref{MasterEq} also holds for the phase accumulation process of photons under one-body losses.

For a given $\phi$-dependent output density matrix $\rho_{out}(\phi)$, the ultimate measurement precision for $\phi$ is imposed by the quantum Cram\'{e}r-Rao bound (QCRB),
\begin{equation}\label{QCRB0}
\Delta \phi \ge \Delta \phi_{min}\equiv \frac{1}{\sqrt{N_m F_{Q}}},
\end{equation}
where the quantum Fisher information (QFI)
\begin{equation}\label{QFI0}
F_{Q}= \textrm{Tr}[L_{\rho}(\rho{'})\rho L_{\rho}(\rho{'})],
\end{equation}
with $\rho'=d\rho(\phi)/d\phi$ and the symmetric logarithmic derivative $L_{\rho}(\rho')$.
To derive the measurement precision $\Delta \phi$, one has to calculate the QFI of the output state to be observed.
Expressing the density matrix in a diagonal form, $\rho_{out} (\phi) = \sum_{j} p_j \left|\psi_j\rangle \langle\psi_j\right|$, the $L_{\rho}(\rho')$ reads as
\begin{equation}\label{SLD}
L_{\rho}(\rho')=\sum_{j,k;p_k+p_j\neq0}\frac{2}{p_k+p_j} \left\langle \psi_k|\rho_{out}'|\psi_j \right\rangle \left|\psi_k \rangle\langle \psi_j\right|,
\end{equation}
and so that the QFI can be given as
\begin{equation}\label{QFI1}
F_{Q}=\sum_{j,k;p_k+p_j\neq0}\frac{2}{p_k+p_j}\left|\langle \psi_k|\rho_{out}'|\psi_j \rangle\right|^2.
\end{equation}

\begin{figure}[htb]
\includegraphics[width=\columnwidth]{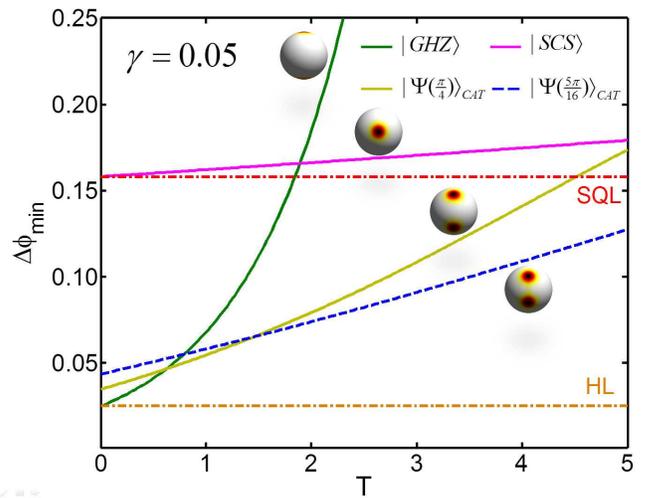}
\caption{(Color online) Ultimate measurement precision $\Delta\phi_{min}$ versus phase accumulation time $T$.
The blue-dashed and yellow-solid lines correspond to the measurement precisions achieved by the spin cat states $|\Psi(\frac{5\pi}{16})\rangle_{\textrm{CAT}}$ and $|\Psi(\frac{\pi}{4})\rangle_{\textrm{CAT}}$, respectively.
The green and pink solid lines correspond to the measurement precisions obtained by the GHZ state ($|\Psi(0)\rangle_{\textrm{CAT}}$) and the SCS ($|\Psi(\frac{\pi}{2})\rangle_{\textrm{CAT}}$), respectively.
The Husimi distributions for these four input states are shown on the generalized Bloch spheres.
The red and orange dash-dotted lines label the SQL and the HL, respectively.
Here, the damping rate $\gamma$ is chosen as $0.05$.}
\label{Fig3}
\end{figure}

Obviously, the measurement precision $\Delta\phi$ is determined by the state $\rho(\phi)$ to be observed.
Actually, due to the accumulated phase $\phi=\delta T$ is a function of the time $T$, the measurement precision $\Delta\phi$ also depends on the time $T$.
In Fig.~\ref{Fig3}, for the input states $|\Psi(\theta)\rangle_{\textrm{CAT}}$ of $\theta=\{0,\pi/4, 5\pi/16, \pi/2\}$, we show how the ultimate measurement precision $\Delta\phi_{min}$ varies with $T$ for $\gamma=0.05$.
Our results show that the $\Delta\phi_{min}$ achieved by the GHZ state ($\theta=0$) increases rapidly with $T$, while the ones achieved by other three states increases slowly.
In comparison with the GHZ state, the other three states are more robust against particle losses.
Up to a modest $T$, the measurement precision achieved by the two spin cat states ($\theta=\pi/4$ and $5\pi/16$) still beat the SQL and are much better than the ones achieved by the GHZ state and the SCS ($\theta=\pi/2$).
This indicates that, for a modest $T$, spin cat states are excellent candidates for implementing dissipative quantum metrology.

Usually, $F_{Q}$ has to be calculated numerically. To simplify the calculation process, an analytical upper bound $\tilde{F}_{Q}$ to $F_{Q}$ for an arbitrary input state $\left|\Psi\right\rangle^{\textrm{in}}=\sum_{k=-J}^{J} C_{m}\left|J,m\right\rangle$ evolving under one-body loss has been given in Refs.~\cite{PhysRevLett.102.040403,PhysRevA.80.013825}.
For different input states $\left|\Psi(\theta)\right\rangle^{\textrm{in}}_{\textrm{CAT}}$ and different values of $\eta$, we calculate the values of $\tilde{F}_{Q}$ and $F_{Q}$. Our results show that the values of $\tilde{F}_{Q}$ are very close to the ones of $F_Q$.
This means that one can use $\tilde{F}_{Q}$ instead of $F_{Q}$ itself to derive the QCRB \eqref{QCRB0}.

To find the optimal input spin cat state $|\Psi(\theta)\rangle_{\textrm{CAT}}$, we calculate the ultimate measurement precision $\Delta\phi_{min}$ for all possible $\theta$.
In Fig.~\ref{Fig2}, we show how $\Delta\phi_{min}$ varies with $\theta$ for different values of $\eta$.
In the absence of particle losses ($\eta=1$), the measurement precision achieved by the GHZ state is better than other ones.
However, in the presence of particle losses ($\eta<1$), the GHZ state becomes fragile and its achievable measurement precision is not the best one.
For all input states, the measurement precision becomes worse and worse when the amount of particle losses becomes larger and larger.
With modest amount of particle losses, most of the input states can still achieve high precision beyond the SQL.
The best optimal measurement precisions (labeled by triangles) and their corresponding input states sensitively depend on the amount of particle losses.
The optimal input state changes from a GHZ state to a spin cat state when the amount of particle losses increases.
For $\eta=\{0.975, 0.95, 0.925, 0.9, 0.8\}$, the optimal input states are the spin cat states of $2\theta \approx \{0.24\pi, 0.49\pi, 0.58\pi, 0.63\pi, 0.73\pi\}$.
In particular, up to a relatively large amount of particle losses ($\eta=0.8$), although the measurement precision achieved by the GHZ state dramatically deteriorate, the measurement precision achieved by the optimal spin cat states can still beat the SQL.
Instead of the GHZ state with maximum entanglement, the spin cat states with moderate $\theta$ are better candidates for implementing precision measurements beyond the SQL.
This means that, in the presence of particle losses, although the GHZ state is fragile, the spin cat states are much more robust for precision metrology.

\begin{figure}[htb]
\centering
\includegraphics[width=\columnwidth]{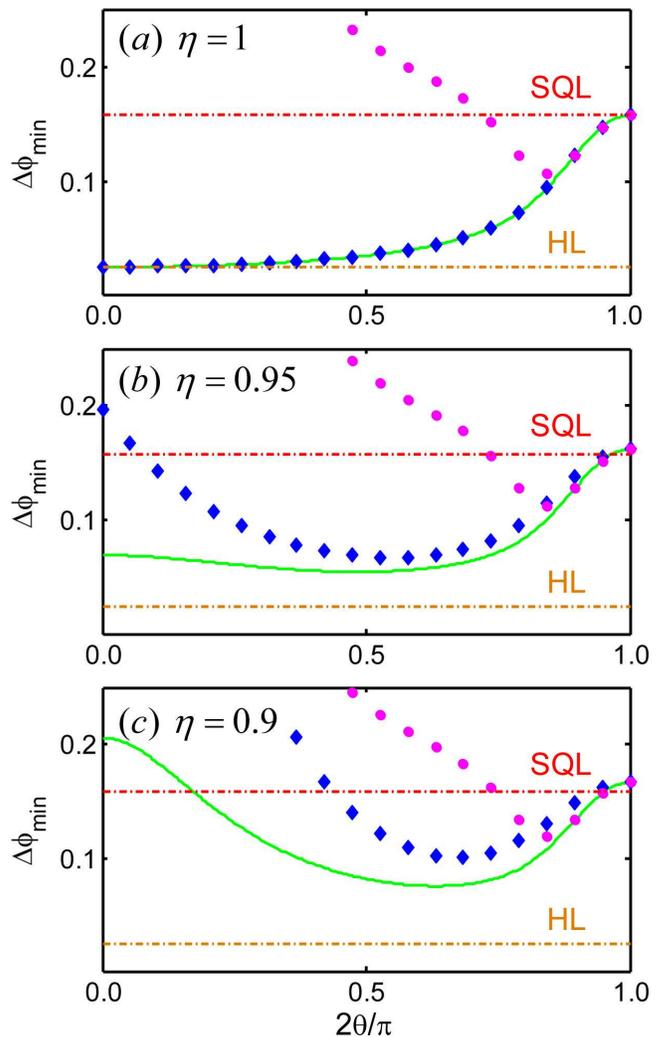}
\caption{(Color online) Phase measurement precision achieved by measuring the parity $\hat{\Pi}_{b}$ (blue diamonds) and by measuring the half population difference $\hat{J}_{z}$ (pink dots) for different amounts of particle losses $(1-\eta)$ with the input spin cat states corresponding to $0 \le \theta \le \pi/2$.
The green curves denote the QCRB.
The red and orange dash-dotted lines correspond to the SQL and the HL, respectively.}
\label{Fig4}
\end{figure}

To extract the phase from the output state, similar to the single-particle Ramsey interferometry, a ${\pi \over 2}$-pulse is applied to the output state and then a suitable observable $\hat{O}$ is observed.
The reduced density matrix for the final state reads as,
\begin{equation}\label{rho_rec}
\rho^{f}=\hat{U}_{\pi/2}^{X\dagger} \rho_{out}(\phi) \hat{U}_{\pi/2}^{X},
\end{equation}
where the unitary operator $\hat{U}_{\pi/2}^{X}=\exp\left[{i\pi \over 2}\hat{J}_{x}\right] =\exp\left[{i \pi \over 4} \left(\hat{a}^{\dagger}\hat{b} +\hat{a}\hat{b}^{\dagger}\right)\right]$.
Therefore, for $N_m$ times of measurements, the phase uncertainty is given as
\begin{equation}\label{EPF}
\Delta\phi=\frac{\Delta\hat{O}}{\sqrt{N_m}|\partial\langle \hat{O}\rangle/\partial\phi|}
\end{equation}
with $\langle\hat{O}\rangle= \textrm{Tr}\left[\hat{O}\rho^{f}\right]$, $\langle\hat{O}^2\rangle= \textrm{Tr}\left[\hat{O}^2\rho^{f}\right]$ and $\Delta\hat{O}=\sqrt{\langle\hat{O}^2\rangle -\langle\hat{O}\rangle^2}$.
Therefore, in a realistic measurement, the measurement precision depends on the input state, the phase accumulation process and the measured observable.
Here, we discuss two typical observables:
the parity $\hat{\Pi}_{b}=\exp(i\pi\hat{b}^{\dagger}\hat{b})$ for mode $b$ and
the half population difference $\hat{J}_{z}=(\hat{b}^{\dagger} \hat{b}-\hat{a}^{\dagger} \hat{a})/2$.
For different amounts of particle losses, according to the formulae~\eqref{EPF}, we calculate the best measurement precision achieved by different input states (see Fig.~\ref{Fig4}).
For the non-dissipative case ($\eta=1$), the measurement of $\hat{\Pi}_{b}$ is optimal for all input spin cat states and the achieved measurement precision (blue diamonds) is completely consistent with the QCRB (green curves).
For dissipative cases ($\eta < 1$), although the precision achieved by measuring $\hat{\Pi}_{b}$ is a bit worse than the QCRB, it still shows similar tendency of the QCRB.
However, for both non-dissipative and dissipative cases, if and only if the input states are close to SCS, the precision achieved by measuring $\hat{J}_{z}$ (pink dots) is well consistent with the QCBR.
Comparing with the $\hat{J}_{z}$-measurement, the parity measurement is more suitable to beat the SQL under particle losses.
Similar to the precisions derived from the QCBR, the ones obtained from the parity measurements also show that the input spin cat states are of excellent robustness against particle losses and the achieved measurement precision can still be much beyond the SQL.

Now, as an example, we discuss potential application of our scheme in realistic phase measurements via atomic Bose-Einstein condensates.
In a recent experiment~\cite{Strobel2014}, enhanced phase measurement has been demonstrated by employing non-Gaussian entangled states of an atomic Bose-Josephson system~\cite{Gross2010,Riedel2010,2014arXiv1405.6022M,CLee2012}.
By using the nonlinear Kerr effects due to atomic collisions, spin cat states can be generated in the Bose-Josephson systems via dynamical evolution~\cite{Sanders1989,Strobel2014} or ground state preparation~\cite{Ho2004,Lee2006,PhysRevLett.102.070401}.
In particular, the self-trapped ground states for symmetric Bose-Josephson systems with negative nonlinearity~\cite{Ho2004,PhysRevLett.102.070401} are very close to the spin cat states~\eqref{CAT}.
The nonlinearity can be controlled by tuning s-wave scattering lengths via Feshbach resonances~\cite{Gross2010} or adjusting spatial overlaps via spin-dependent forces~\cite{Riedel2010}.
Once the desired input states are prepared, the inter-mode coupling and the nonlinearity are turned off and the system evolves under a field-free evolution for phase accumulation.
During the phase accumulation, due to the collisions between Bose condensed atoms and residual thermal atoms~\cite{Strobel2014, Ockeloen2013}, the system obeys the master equation~\eqref{MasterEq}.
At last, the parity/population measurement of the final state can be accomplished via the available techniques of counting atoms~\cite{Hume2013}.

In summary, we have presented a robust scheme for implementing dissipative quantum metrology with spin cat states.
We find that the input spin cat states are of excellent robustness against particle losses and may still achieve high-precision measurements beyond the SQL.
Furthermore, we also investigate the cases of two-body particle losses as well as correlated dephasing and the results also indicate that the input spin cat states are robust~\cite{TBP}.
In realistic measurements based upon our scheme, our analysis shows that the parity measurement is more suitable for yielding high precision beyond the SQL.
It is possible to utilize our scheme for realistic phase measurements with dissipative many-body systems of Bose condensed atoms.

The authors thank Markus K. Oberthaler for his valuable suggestion. This work is supported by the National Basic Research Program of China (NBRPC) under Grant No. 2012CB821305, the National Natural Science Foundation of China (NNSFC) under Grants No. 11374375, and the PhD Programs Foundation of Ministry of Education of China under Grant No. 20120171110022.



\begin{thebibliography}{99}

\bibitem{Giovannetti2004}
G. Vittorio, L. Seth, and M. Lorenzo, Science {\bf 306}, 1330 (2004).

\bibitem{Giovannetti2011}
G. Vittorio, L. Seth, and M. Lorenzo, Nat. Photonics {\bf 5}, 222 (2011).

\bibitem{LeeBookChapter2014}
J. Huang, H. Zhong, S. Wu, and C. Lee, Annual Review of Cold Atoms and Molecules, Volume 2, Chapter 7 \emph{``QUANTUM METROLOGY WITH COLD ATOMS''}, pp. 365-415 (2014).

\bibitem{Georgescu2014}
I. Georgescu, Nat. Phys. {\bf 10}, 474 (2014).

\bibitem{Giovannetti2006}
V. Giovannetti, S. Lloyd, and L. Maccone, Phys. Rev. Lett. {\bf 96}, 010401 (2006).

\bibitem{Lee2006}
C. Lee, Phys. Rev. Lett. {\bf 97}, 150402 (2006).

\bibitem{Boixo2008}
S. Boixo, A. Datta, M. J. Davis, S. T. Flammia, A. Shaji, and C. M. Caves, Phys. Rev. Lett. {\bf 101}, 040403 (2008).

\bibitem{Pezze2009}
L. Pezze and A. Smerzi, Phys. Rev. Lett. {\bf 102}, 100401 (2009).

\bibitem{Liu2011}
Y. C. Liu, Z. F. Xu, G. R. Jin, and L. You, Phys. Rev. Lett. {\bf 107}, 013601 (2011).

\bibitem{Swallows25022011}
M. D. Swallows, M. Bishof, Y. Lin, S. Blatt, M. J. Martin, A. M. Rey, and J. Ye, Science {\bf 331}, 1043 (2011).

\bibitem{Martin09082013}
M. J. Martin, M. Bishof, M. D. Swallows, X. Zhang, C. Benko, J. von Stecher, A. V. Gorshkov, A. M. Rey, and J. Ye, Science {\bf 341}, 632 (2013).

\bibitem{PhysRevLett.92.230801}
A. Andr\'{e}, A. S. S{\o}rensen, and M. D. Lukin, Phys. Rev. Lett. {\bf 92}, 230801 (2004).

\bibitem{PhysRevLett.110.093602}
H. M\"{u}ntinga, H. Ahlers, M. Krutzik, A. Wenzlawski, S. Arnold, D. Becker, K. Bongs, H. Dittus, H. Duncker, N. Gaaloul, C. Gherasim, E. Giese, C. Grzeschik, T. W. H\"{a}nsch, O. Hellmig, W. Herr, S. Herrmann, E. Kajari, S. Kleinert, C. L\"{a}mmerzahl, W. Lewoczko-Adamczyk, J. Malcolm, N. Meyer, R. Nolte, A. Peters, M. Popp, J. Reichel, A. Roura, J. Rudolph, M. Schiemangk, M. Schneider, S. T. Seidel, K. Sengstock, V. Tamma, T. Valenzuela, A. Vogel, R. Walser, T. Wendrich, P. Windpassinger, W. Zeller, T. van Zoest, W. Ertmer, W. P. Schleich, and E. M. Rasel, Phys. Rev. Lett. {\bf 110}, 093602 (2013).

\bibitem{PhysRevD.65.022002}
H. J. Kimble, Y. Levin, A. B. Matsko, K. S. Thorne, and S. P. Vyatchanin, Phys. Rev. D {\bf 65}, 022002 (2011).

\bibitem{PhysRevLett.104.133601}
W. Wasilewski, K. Jensen, H. Krauter, J. J. Renema, M. V. Balabas, and E. S. Polzik, Phys. Rev. Lett. {\bf 104}, 133601 (2010).

\bibitem{PhysRevA.82.022330}
S. Simmons, J. A. Jones, S. D. Karlen, A. Ardavan, and J. J. L. Morton, Phys. Rev. A {\bf 82}, 022330 (2010).

\bibitem{Ng2013}
H. T. Ng, Phys. Rev. A {\bf 87}, 043602 (2013).

\bibitem{Gross2010}
G. Gross, T. Zibold, E. Nicklas, J. Est\`{e}ve, and M. K. Oberthaler, Nature (London) {\bf 464}, 1165 (2010).

\bibitem{Riedel2010}
M. F. Riedel, P. B\"{o}hi, Y. Li, T. W. H\"{a}nsch, A. Sinatra, and P. Treutlein, Nature (London) {\bf 464}, 1170 (2010).

\bibitem{Grond2011}
J. Grond, U. Hohenester, J. Schmiedmayer, and A. Smerzi, Phys. Rev. A {\bf 84}, 023619 (2011).

\bibitem{Lucke2011}
B. L\"{u}cke, M. Scherer, J. Kruse, L. Pezz\'{e}, F. Deuretzbacher, P. Hyllus, O. Topic, J. Peise, W. Ertmer, J. Arlt, L. Santos, A. Smerzi, and C. Klempt, Science {\bf 334}, 773 (2011).

\bibitem{Berrada2013}
T. Berrada, S. van Frank, R. B\"{u}cker, T. Schumm, J. -F. Schaff, and J. Schmiedmayer, Nat. Communication {\bf 4}, 2077 (2013).

\bibitem{2014arXiv1405.6022M}
W. Muessel, H. Strobel, D. Linnemann, D. B. Hume, and M. K. Oberthaler, arXiv e-prints (2014), 1405.6022.

\bibitem{UedaPRA}
M. Kitagawa and M. Ueda. Phys. Rev. A {\bf 47}, 5138 (1993).

\bibitem{Strobel2014}
H. Strobel, W. Muessel, D. Linnemann, T. Zibold, D. B. Hume, L. Pezz\`{e}, A. Smerzi, M. K. Oberthaler, Science {\bf 345}, 6195 (2014).

\bibitem{PhysRevLett.102.040403}
U. Dorner, R. Demkowicz-Dobrzanski, B. J. Smith, J. S. Lundeen, W. Wasilewski, K. Banaszek, and I. A. Walmsley, Phys. Rev. Lett. {\bf 102}, 040403 (2009).

\bibitem{PhysRevA.80.013825}
R. Demkowicz-Dobrzanski, U. Dorner, B. J. Smith, J. S. Lundeen, W. Wasilewski, K. Banaszek, and I. A. Walmsley, Phys. Rev. A {\bf 80}, 013825 (2009).

\bibitem{Joo2011}
J. Joo, W. J. Munro, and T. P. Spiller, Phys. Rev. Lett. {\bf 107}, 083601 (2011).

\bibitem{PhysRevA.83.043620}
Y. Hao and Q. Gu, Phys. Rev. A {\bf 83}, 043620 (2011).

\bibitem{Ockeloen2013}
C. F. Ockeloen, R. Schmied, M. F. Riedel, and P. Treutlein, Phys. Rev. Lett. {\bf 111}, 143001 (2013).

\bibitem{Pawlowski2013}
K. Paw{\l}owski, D. Spehner, A. Minguzzi, and G. Ferrini, Phys. Rev. A {\bf 88}, 013606 (2013).

\bibitem{PhysRevLett.100.210401}
Y. Li, Y. Castin, and A. Sinatra, Phys. Rev. Lett. {\bf 100}, 210401 (2008).

\bibitem{Escher2011}
B. M. Escher, R. L. de Matos, and L. Davidovich, Nat. Phys.  {\bf 7}, 406 (2011).

\bibitem{Alipour2014}
S. Alipour, M. Mehboudi, and A. T. Rezakhani, Phys. Rev. Lett. {\bf 112}, 120405 (2014).

\bibitem{Ho2004}
T. -L. Ho, and C. V. Ciobanu, Journal of Low Temperature Physics, {\bf 135}, 257 (2004).

\bibitem{Lau2014}
H. Lau, Z. Dutton, T. Wang, and C. Simon, Phys. Rev. Lett. {\bf 113}, 090401 (2014).

\bibitem{Hume2013}
D. Hume, I. Stroescu, M. Joos, W. Muessel, H. Strobel, and M. Oberthaler, Phys. Rev. Lett. {\bf 111}, 253001 (2013).

\bibitem{Agarwal1997}
G. S. Agarwal, R. R. Puri, and R. P. Singh, Phys. Rev. A {\bf 56}, 2249 (1997).

\bibitem{Gerry1997}
C. C. Gerry and R. Grobe, Phys. Rev. A {\bf 56}, 2390 (1997).

\bibitem{Wang2000}
X. Wang, B. C. Sanders, and S. Pan, J. Phys. A: Math. Gen. {\bf 33}, 7451 (2000).

\bibitem{Sanders2014}
B. C. Sanders, arXiv e-prints (2014), 1405.5026.

\bibitem{Sanders1989}
B. C. Sanders, Phys. Rev. A {\bf 40}, 2417 (1989).

\bibitem{PhysRevLett.102.070401}
C. Lee, Phys. Rev. Lett. {\bf 102}, 070401 (2009).

\bibitem{Gerry1998}
C. C. Gerry and R. Grobe, Phys. Rev. A {\bf 57}, 2247 (1998).

\bibitem{Dooley2013}
S. Dooley, F. McCrossan, D. Harland, M. J. Everitt, and T. P. Spiller, Phys. Rev. A {\bf 87}, 052323 (2013).

\bibitem{McConnell2013}
R. McConnell, H. Zhang, S. $\acute{C}$uk, J. Hu, M. H. Schleier-Smith, and V. Vuleti$\acute{c}$, Phys. Rev. A {\bf 88}, 063602 (2013).

\bibitem{Signoles2014}
A. Signoles, A. Facon, D. Grosso, I. Dotsenko, S. Haroche, J. -M. Raimond, M. Brune, and S. Gleyzes, Nat. Phys. {\bf 10}, 715 (2014).

\bibitem{PhysRevA.58.R50}
J. Ruostekoski and D. F. Walls, Phys. Rev. A {\bf 58}, R50 (1998).

\bibitem{pawlowski2010}
K. Paw{\l}owski and K. Rz\c{a}\.{z}ewski, Phys. Rev. A {\bf 81}, 013620 (2010).

\bibitem{CLee2012}
C. Lee, J. Huang, H. Deng, H. Dai, and J. Xu, Frontiers of Physics {\bf 7}, 109 (2012).

\bibitem{TBP}
We have calculated the ultimate measurement precisions for the input spin cat states $\left|\Psi(\theta)\right\rangle_{\textrm{CAT}}$ under two-body particle losses and correlated dephasing. For an initial state $\left|\Psi(\theta)\right\rangle_{\textrm{CAT}}$ with a fixed damping rate of two-body losses, the amounts of particle losses during the phase accumulation decrease with $\theta$ and this leads to more rapid reduction of the measurement precisions for smaller $\theta$. The input state $\left|\Psi(\theta)\right\rangle_{\textrm{CAT}}$ with moderate $\theta$ can still achieve robust high-precision measurement beyond the SQL. However, in the case of one-body losses, for a fixed damping rate, the amounts of particle losses during the phase accumulation only depends on the evolution time $T$ but not depends on $\theta$.
In the case of correlated dephasing, only the coherence of relative phases is destroyed in the phase accumulation process, and, for a fixed dephasing rate, the ultimate precisions versus the accumulation time shows similar tendency of the case of one-body losses.
The details of these results will be published elsewhere.

\end{thebibliography}
\end{document}